\documentclass[10pt]{article}
\usepackage[left=1in,right=1in,top=1in,bottom=1in]{geometry}
\usepackage{url,lineno,microtype,subcaption,color}
\definecolor{niceblue}{RGB}{0,0,100}
\usepackage[colorlinks=true,linkcolor=niceblue,citecolor=niceblue,urlcolor=blue]{hyperref}
\usepackage[onehalfspacing]{setspace}
\usepackage{arydshln}
\usepackage{upgreek}
\usepackage{graphicx}
\usepackage{amsmath}

\usepackage[maxbibnames=99,citestyle=authoryear,style=authoryear,backend=bibtex,hyperref]{biblatex}
\usepackage{hyperref}

\renewbibmacro*{pageref}{%
 \printtext[parens]{%
  \Acrobatmenu{GoBack}{Go back}%
 }%
}

\begin{document}
%\onecolumn
%\firstpage{1}

%% Front matter %%

%\title[Velocity of Money]{The Transfer Velocity of Money}

%\author[\firstAuthorLast ]{\Authors} %This field will be automatically populated
%\address{} %This field will be automatically populated
%\correspondance{} %This field will be automatically populated

%\extraAuth{}

%\maketitle

% Title Page
	\begin{titlepage}
	\title{\LARGE \bf Holding Periods: Measuring the Inverse of Money Velocity from Transaction Records}
	\author{Carolina Mattsson\footnote{CENTAI Institute, Corso Inghilterra 3, 10138, Turin, Italy. carolina.mattsson@centai.eu.}\\ CENTAI Institute \and Allison Luedtke\footnote{Department of Economics, St. Olaf College. 1520 St. Olaf Avenue, Northfield, MN 55057}\\St. Olaf College \and Frank Takes\footnote{Leiden Institute of Advanced Computer Science, Leiden University, Leiden, Netherlands. f.w.takes@liacs.leidenuniv.nl.}\\Leiden Institute of Advanced\\ Computer Science \vspace{5mm}}
	\date{\today}
	\maketitle
	
	%\bigskip
	\begin{abstract}
		\normalsize 
        This paper defines the average holding period of money and proposes a methodology for fully disaggregated measurement. Our measure is shown to be the inverse of the transfer velocity of money under stationary conditions, which is implicitly assumed in conventional aggregate measurement. Our methodology does not require stationarity. We leverage a recent computational technique to extract empirical holding periods from micro-level transaction data as recorded by real-world payment systems. This enables novel empirical analyses of money velocity under non-stationarity conditions. We illustrate several such analyses on Sarafu, a small digital community currency in Kenya, where transaction data is available from 25 January 2020 to 15 June 2021. Our measure implies faster circulation than does the aggregate transfer velocity; we can say that 58\% of Sarafu was effectively static. We also disaggregate by geography to study the heterogeneous impact of economic disruptions related to the COVID-19 pandemic on Sarafu. Finally, we consider an ad-hoc currency management operation that took place in October 2020. Measuring the average holding period of money makes it possible to track money velocity during ongoing monetary interventions and on other important occasions when conditions are not stationary.
	\end{abstract}
	%\vspace{3cm}
	\setcounter{page}{0}
	\thispagestyle{empty}

\end{titlepage}
\pagebreak \newpage

%% Main text %%
%\doublespacing
\section{Introduction}

The rate at which money changes hands is an important macroeconomic indicator and plays a key role in determining inflation \parencite{benati_money_2020, benati_new_2023}. Money tends to move faster between people when the economy is doing well and slower when the economy is doing poorly~\parencite{leao_why_2005}. However, conventional measures for the velocity of money are typically constructed as the ratio of two large macroeconomic aggregates. For example, on the Federal Reserve Economic Data (FRED) online database, the Velocity of M2 (FRED ID: \texttt{M2V}) is calculated as the ratio of quarterly nominal gross domestic product to \texttt{M2}, a measure of the aggregate money supply~\parencite{federal_reserve_bank_of_st_louis_velocity_2022}. This methodology reflects classic formulations of money velocity as the term that balances a ``quantity equation'' of money. Quantity equations are constructed from aggregate accounting identities under theoretical frameworks that have long been of interest to monetary economists. Fisher favored a quantity equation constructed from the sum total of transactions in an economy, Friedman adopts one that represents money as an asset, and the income form of the quantity equation that underpins \texttt{M2V} is something of a compromise~\parencite{friedman_theoretical_1970}. Some decades later, Leontief advanced the earlier, more mechanistic, transactions perspective arguing that the use of income-based quantity equations ``severely underestimates the actual speed [of money] and even more so the speed of circulation of the goods and services money is carrying''~\parencite[][p. 231]{leontief_money-flow_1993}.

Recent developments heighten the appeal of the mechanistic perspectives. Transactions aggregates have historically been difficult to measure~\parencite{cramer_volume_1986} but modern payment infrastructure is largely digital. Money changing hands can be nearly perfectly observed in digitally recorded transaction data. The US Federal Reserve, Banca d'Italia, and private banks have used micro-level transaction records from retail payment processors to provide timely estimates of macroeconomic aggregates~\parencite{aladangady_transaction_2022,aprigliano_using_2019,buda_national_2022}. National payment systems are now being used to produce experimental data series on payment flows in the UK (inter-industry) and in Brazil (inter-firm)~\parencite{hotte_national_2024,silva_modeling_2022}. For a given payment system, precise record-keeping allows for the measurement of an empirical quantity: the {\it transfer velocity of money}. This value is conventionally calculated as the ratio between the total transfer volume in a given time frame and the average total balance held by users of the payment system~\parencite{mbiti_home_2013}. Although this metric coexists uneasily with the classic formulations of money velocity, the quality of the data recorded by digital payment infrastructure is highly compelling. Given comprehensive transaction records, it has become computationally feasible to measure the transfer velocity of money down to the level of individual accounts~\parencite[MicroVelocity:][]{campajola_microvelocity_2022,collibus_microvelocity_2025}. 

This work re-examines existing formulations of money velocity, aiming to make it also conceptually feasible to measure money velocity down to the level of individual accounts. Already in aggregate, the dynamics of money velocity have been found to be stochastic and possibly chaotic~\parencite{serletis_random_1995} or with inconsistent behavior across different dynamical regimes~\parencite{ardakani_dynamics_2023}. Moreover, the economics literature has long recognized that money velocity differs across payment systems, between sectors, and among individuals~\parencite{cramer_volume_1986,leontief_money-flow_1993,mbiti_home_2013}. Direct, disaggregated, empirical measurement in this setting raises complications that existing frameworks and definitions are unable to address. Namely, instability in the denominator (the balance) on the timescale required to measure the numerator (the transaction volume). The issue becomes especially clear at finer resolutions. Consider conventional measurement of the transfer velocity for an individual account: each transaction that contributes $y$ to the transfer volume spent by the account will also reduce the balance of that account by $y$. Whatever is the time frame considered, the ostensible denominator will have fluctuated by at least as much as the numerator.

Prior efforts at disaggregation, in theory and empirically, avoid this complication by implicitly or explicitly assuming stationarity. \textcite{leontief_money-flow_1993} express the Fisher equation for the velocity of money in the context of input-output analysis, complemented with monetary stock-flow data. The ergodic state of a Markov chain, explicitly a stationary condition, is used to infer the sector-specific velocities of money that would give rise to monetary flows as measured between sectors given also the share of money held by those sectors (i.e., their balance). The aggregate velocity of money would be the average over the sector-specific velocities weighted by the stationary balance of each sector. \textcite{cramer_volume_1986} and, more recently, \textcite{mbiti_mobile_2015} estimate the effect of improved payment technologies on money demand. Both note that modern currency systems are made up of many payment systems with widely different transfer velocities, and, in the computation of an overall average, implicitly assume that the relative total balances, as measured, can be seen as fixed. Considering goods rather than money, \textcite{antras_measuring_2025} propose a methodology for measuring the average period of production. By assuming stationary conditions, the authors arrive at an empirical computation involving the ratio of inventories to the cost of goods sold. This is directly analogous to conventional measurement of the transfer velocity of money, in inverse. Indeed, one can also consider ``holding periods'' of money except that money is not consumed at the end of a period. Funds enter an account, are held there for some period, and then are transferred out \textit{to another account}. The next period begins just as the prior period ends. This perspective has been used to define equations for the velocity of money in the context of simulated transaction processes, where a stationary distribution of holding periods can be derived~\parencite{wang_circulation_2003, wang_prospects_2005, kanazawa_kinetic_2018}. Taking this approach, \textcite{campajola_microvelocity_2022} develop empirical measurement of the transfer velocity of money at the level of individual accounts, retaining implicitly the assumption of stationary conditions.

% Contribution in terms of METHOD
This paper starts by precisely defining the average holding period of money. Our definition reduces to the inverse of the transfer velocity of money under stationary conditions, but it also generalizes to non-stationary conditions. We then provide a methodology for measuring the average holding period of money for a real-world payment system given micro-level transaction data as recorded by that payment system. The transaction records are used to construct a weighted, directed, temporal network describing transfers among accounts within the system. Holding periods are extracted using an existing computational technique that traces funds through the temporal network~\parencite{mattsson_trajectories_2021}. We compute the average holding period of money as the weighted average of the holding periods observed to completion within a given time frame. This value can be computed for the full system, and for any subset of accounts. The inverse of this measure has the units of a money velocity and can be compared to the conventionally measured aggregate transfer velocity. In this way, our measure and our measurement methodology enable novel empirical analysis of money velocity under non-stationarity conditions. 

% Contribution in terms of DATA 
We illustrate our approach on data from Sarafu, a small digital community currency in Kenya, where transaction data is available from 25 January 2020 to 15 June 2021. The published data set includes around 400,000 transactions among what grew to be approximately 40,000 accounts, plus limited information about the account holders~\parencite{ruddick_sarafu_2021}. The Sarafu system has several unique features and this time frame invites us to question the assumption of stationarity. Community currencies are much smaller than national currencies and are known to behave differently~\parencite{stodder_complementary_2009,stodder_macro-stability_2016,zeller_economic_2020}. Rates of use tend to be counter-cyclical and Kenya experienced substantial economic disruption at the start of the COVID-19 pandemic~\parencite{fews_net_kenya_2020}. Sarafu became part of documented humanitarian and pandemic-response efforts~\parencite{ussher_complementary_2021}. Moreover, there are documented instances of currency interventions administered in the management of Sarafu over this time frame~\parencite{mattsson_sarafu_2022}.

% Ovrerall Contribution in terms of results: look it works!
Applying our measure to the Sarafu data shows that, indeed, circulation within the Sarafu system was non-stationary. Overall, we find that the average holding period of Sarafu implies faster circulation than does the aggregate transfer velocity. The discrepancy is such that only 42\% of the issued balance of Sarafu would be needed to produce the observed volume of transfers if circulation were stationary with the observed average holding period; in this sense, 58\% of the total Sarafu balance was effectively static. This is reflected also in the empirical distribution of completed holding periods---some tokens of Sarafu were held for minutes, others for months. Notably, strong heterogeneity in velocity has also been found for mobile money payment systems and cryptocurrencies~\parencite{mbiti_home_2013,campajola_microvelocity_2022,collibus_microvelocity_2025}. The potential for non-stationary dynamics within payment and currency systems is underexplored even as non-stationarity has been incorporated into network representations of the relationships among currency systems, such as in the foreign exchange market~\parencite{chen_large_2024} and cryptocurrency market~\parencite{elbahrawy_evolutionary_2017,guo_time-varying_2024}. Rising interest in digital financial services, real-time gross settlement, and central bank digital currencies makes the topic more relevant. In dispensing with the assumption that real-world systems operate under stationary conditions we enable deeper empirical study of emerging payment technologies and of increasingly fragmented currency systems.

The added value of our proposed approach for disaggregated measurement of the average holding period of money is that it allows us to better characterize money velocity during shocks and interventions.

% subresult shocks 
Starting with the first, we empirically study the localized impact of shocks on Sarafu by, over time, tracking the velocity of its circulation within communities of users in urban Nairobi and in rural Kinango Kwale, two areas that both experienced pandemic-related economic disruptions. There were sizeable differences, and change over time, in the velocity of Sarafu in these areas. Our findings are consistent with the prevailing understanding of community currencies in periods of economic disruption~\parencite{stodder_complementary_2009,zeller_economic_2020}. More broadly, there is growing evidence that many macroeconomic measures, such as inflation and unemployment, and important macroeconomic channels such as monetary policy, affect individuals, groups, and locales differently~\parencite{argente_cost_2021,gornemann_doves_2021,bartscher_monetary_2021,goodman-bacon_myth_2021}. Our methodological advance contributes to this growing literature in that we use new data and empirical techniques to identify and analyze heterogeneity in a measure of money velocity.

% subresult interventions
Second, we show that our methodology offers a way to monitor money velocity while interventions are taking place. The Sarafu system is managed by Grassroots Economics Foundation, a Kenyan nonprofit organization, and a surge in signups during the early pandemic complicated this task over this particular time frame. In October 2020, Grassroots Economics targeted inactive accounts with an operation that removed 27\% of Sarafu from circulation in less than two days~\parencite{mattsson_sarafu_2022}. We characterize this currency operation and compute the average holding period of Sarafu as the operation took place. Interestingly, we see no response among spenders in Nairobi. In Kinango Kwale, transaction volumes rose without a rise in velocity as ``old'' Sarafu re-entered active circulation. Our generalized monetary indicators remain properly defined when the monetary base expands or contracts suddenly, offering a major advantage in times of crisis. Monitoring the money supply is an important prerequisite for conducting sound monetary policy, and our methodology offers a way to do so using the digital records of the system itself. 

The remainder of this article is organized as follows: Section~\ref{sec:measure} introduces the average holding period of money and formally defines its relation to the transfer velocity of money. In Section~\ref{sec:method} we detail our methodology for measuring this value from transaction data as recorded by real-world payment systems. Section~\ref{sec:data} describes the Sarafu system, our example application. We present novel empirical analyses of this system in Section~\ref{sec:results}. Section~\ref{sec:conclusion} concludes.

\section{Average holding period of money} \label{sec:measure}

The transfer velocity of money is conventionally defined using an identity similar to a quantity equation. The total transfer volume or the total flow of money, $F_T$, in a time window, $t_0 < t < t_1$, is related to the amount of money in circulation, $M$, and its aggregate transfer velocity, $V$. Eqn.~\ref{eqn:flow} describes the familiar relation. We include the length of time $\Delta t = t_1 - t_0$ in our formulation, explicitly, so that both $F_T$ and $M$ are amounts denoted in units of currency. The scale of $V$ can be defined, implicitly, such that $\Delta t = 1$.

\begin{equation}
F_T =  M \cdot \Delta t \cdot V
\label{eqn:flow}
\end{equation}

The transfer velocity can also be defined using the concept of a ``holding period.'' Denoted $\uptau$, holding periods are the durations between when accounts receive and re-transact particular units of money. \cite[][Eqn. 4]{wang_circulation_2003} define $P(\uptau)$ as the probability density for a given unit of money being used in a transaction after an interval of $\uptau$. This could be considered a waiting time distribution with respect to the units of money. Integrating over this distribution is to consider all units of money at a snapshot in time, allowing an expression for the transfer velocity \parencite[][Eqn. 8]{wang_circulation_2003}:

\begin{equation}
V = \int_{0}^{\infty} { P(\uptau) \cdot \frac{1}{\uptau} \cdot \mathrm{d}\uptau}
\text{   where   }
1 = \int_{0}^{\infty} {P(\uptau) \mathrm{d}\uptau}
\label{eqn:velocity}
\end{equation}

The definitions in Eqns.~\ref{eqn:flow} and~\ref{eqn:velocity} both assume that $M$ is fixed, or, at least, that $M(t)$ is well represented by its time-average~\parencite{mbiti_home_2013}. Moreover, to arrive at Eqn.~\ref{eqn:velocity}, \textcite{wang_circulation_2003} consider a stochastic transaction process in its stationary state, where $P(\uptau)$ is independent of $t$. Here we introduce $P(\uptau,t)$ as the non-stationary generalization of $P(\uptau)$. While a waiting time  distribution that changes in time is conceptually difficult to relate to observables of a real system, it proves useful in derivations. Eqn.~\ref{eqn:flow_generalized} formulates a generalized version of the expression for $F_T$ without assuming stationary conditions, that is, where both $M$ and $P(\uptau)$ are functions also of $t$~\parencite[cf.][Eqn. 7]{wang_circulation_2003}. Eqn.~\ref{eqn:flow_dist} uses differential form to express the transfer volume generated by the share of money transacted after a period of exactly $\uptau$ ~\parencite[cf.][Eqn. 6]{wang_circulation_2003}.

\begin{equation}
F_T = \int_{t_0}^{t_1} \int_{0}^{\infty} M(t)  P(\uptau,t) \cdot \frac{1}{\uptau} \cdot \partial \uptau \partial t
\label{eqn:flow_generalized}
\end{equation}

\begin{equation}
F_T(\uptau) =  \int_{t_0}^{t_1} M(t) P(\uptau,t) \cdot \frac{1}{\uptau} \cdot \mathrm{d}t
\label{eqn:flow_dist}
\end{equation}

Notably, $F_T(\uptau)$ is in principle observable for real systems. This would be the distribution of holding periods completed over $t_0 < t < t_1$, that is, prior to the transactions in this time window. Eqn.~\ref{eqn:flow_dist} then lets us define the average period of time that money was held prior to being used in a transaction during the window $t_0 < t < t_1$. This is the average holding period, denoted as $\bar{\uptau}_T$ in Eqn.~\ref{eqn:duration_dist}:

\begin{equation}
\bar{\uptau}_T =  \frac{1}{F_T} \int_{0}^{\infty} F_T(\uptau) \cdot \uptau \cdot \mathrm{d}\uptau
\label{eqn:duration_dist}
\end{equation}

We can now show that the average holding period is the inverse of the aggregate transfer velocity under stationary conditions. Incorporating Eqn.~\ref{eqn:flow_dist} into Eqn.~\ref{eqn:duration_dist} produces Eqn.~\ref{eqn:holding}. When $M(t)$ and $P(\uptau,t)$ are independent of $t$, the double integral simplifies to $M \cdot \Delta t$. Rearranging the simplified expression gives us Eqn.~\ref{eqn:flow2}. Note the parallel with Eqn.~\ref{eqn:flow}, now with $\bar{\uptau}_T^{-1}$ taking the place of $V$. 

\begin{equation}
\bar{\uptau}_T = \frac{1}{F_T} \int_{0}^{\infty} \int_{t_0}^{t_1} M(t) P(\uptau,t) \partial t \partial \uptau
\label{eqn:holding}
\end{equation}

\begin{equation}
F_T = M \cdot \Delta t \cdot \bar{\uptau}_T^{-1}
\label{eqn:flow2}
\end{equation}

Let us call the inverse of the average holding period the non-stationary transfer velocity. This is denoted as $V_T$ in Eqn.~\ref{eqn:velocity_dist} because it is defined over the time window $t_0 < t < t_1$ in the same sense as is $F_T$. The non-stationary transfer velocity $V_T$ will generally differ from the aggregate transfer velocity $V$. 

\begin{equation}
V_T = \bar{\uptau}_T^{-1}
\label{eqn:velocity_dist}
\end{equation}

Unlike the aggregate transfer velocity $V$, the non-stationary transfer velocity $V_T$ does not rest on estimates of $M$. This means we can define the \textit{effective balance} of the system over $t_0 < t < t_1$ as the fixed balance that would satisfy the relationship between the average holding period $\bar{\uptau}_T$ and the total transfer volume $F_T$ under stationarity. The effective balance $M_T$ is defined in Eqn.~\ref{eqn:balance_eff}, drawing from Eqn.~\ref{eqn:holding}. 

\begin{equation}
M_T =  \frac{1}{\Delta t} \int_{0}^{\infty} \int_{t_0}^{t_1} M(t) P(\uptau,t) \partial t \partial \uptau = \frac{1}{\Delta t} \cdot F_T \cdot \bar{\uptau}_T
\label{eqn:balance_eff}
\end{equation}

\section{Measurement methodology} \label{sec:method}

The transfer velocity of money quantifies the speed of circulation for funds within a specific payment system. Payment systems are increasingly digital and digital transaction records offer new opportunities for precise measurement. Section~\ref{sec:method:conventional} describes the conventional approach to measuring the aggregate transfer velocity of funds from micro-level transaction data as recorded by a payment system. Our methodology for measuring the average holding period of funds from transaction data is introduced in Section~\ref{sec:method:inverse}.

\subsection{Measuring the aggregate transfer velocity} \label{sec:method:conventional}

The aggregate transfer velocity $V$ is computed using Eqn.~\ref{eqn:flow}. The total flow of money $F_T$ is a directly measurable quantity and the total balance $M$ that is issued within the system can be estimated. In practice, one considers the time-average balance over the time window $t_0 < t < t_1$~\parencite{mbiti_home_2013}. Eqn.~\ref{eqn:velocity_conv} formulates an expression for the aggregate transfer velocity $V$ using the empirical time-average of $M(t)$. Times $t_0$ and $t_1$ can be selected so that $\Delta t = t_1 - t_0 = 1$ in the time unit also used for reporting $V$.

\begin{equation}
V =  F_T \big/ ( M_{\text{avg}} \cdot \Delta t )
\text{   where   }
M_{\text{avg}} = \int_{T_0}^{T_1} {M(t) \mathrm{d}t} \big/ \Delta t
\label{eqn:velocity_conv}
\end{equation}

Empirical values for $F_T$ and $M(t)$ can be measured from micro-level transaction data. The total flow $F_T$ is the sum of transaction sizes for all transfer transactions that occur in the time window $t_0 < t < t_1$. The total balance $M(t)$ is the sum of the amounts of money held in that moment by all user-facing accounts; this needs to be measured at regular intervals or inferred from records of issuance and dissolution. Note that balances held by provider-facing accounts are generally not included in the total~\parencite{mbiti_home_2013}.

\subsection{Measuring the average holding period} \label{sec:method:inverse}

The average holding period of money $\bar{\uptau}_T$ is computed using Eqn.~\ref{eqn:duration_dist}. The distribution $F_T(\uptau)$ over the period $t_0 < t < t_1$ can be obtained from micro-level transaction data. Specifically, a computational technique called ``trajectory extraction'' can be used to obtain empirical holding periods from transaction data~\parencite{mattsson_trajectories_2021}. Funds are matched from incoming to outgoing transactions, giving the holding periods $\uptau$. The holding periods that end in transfer transactions occurring in the time window $t_0 < t < t_1$ are collected into an empirical distribution; this is $F_T(\uptau)$. Eqn.~\ref{eqn:duration_dist} is evaluated as a weighted average.

\textcite{mattsson_trajectories_2021} details trajectory extraction, a data transformation based in the theory of walk processes on networks. Transactions out of an account are allocated funds from prior transactions into that account. We consider all possible pairs of transactions by selecting the ``well-mixed'' allocation heuristic in the algorithm. This means that an amount is assigned to every possible pair reflecting proportional allocation of funds; this is computationally intensive. Note that simpler heuristics are available for use with large datasets and produce precise estimates in practice~\parencite{collibus_microvelocity_2025}. The outgoing transaction completes the holding period and its timestamp $t$ is also the moment where the observation takes place.

\section{Application: Sarafu payment data} \label{sec:data}

The Sarafu system is a set of small digital community currencies in Kenya operated by Grassroots Economics Foundation, a Kenyan nonprofit organization whose efforts are concentrated in specific areas and aimed at supporting marginalized, food insecure communities. In 2020-21 the system comprised a single community currency (Sarafu) under centralized management procedures.\footnote{Earlier iterations of the Sarafu system favored decentralized currency management procedures, as do subsequent systems in place since April 2022 and July 2023. The highly centralized system studied here was a temporary stop-gap technology; it remained in place longer than intended as Grassroots Economics adjusted priorities during the COVID-19 pandemic.} Payments in Sarafu were made via a mobile phone interface. One token unit of Sarafu was roughly equivalent in value to a Kenyan shilling but the system itself was completely separate from the national currency. Community currencies behave differently than do national currencies~\parencite{stodder_complementary_2009,stodder_macro-stability_2016,zeller_economic_2020}. Here we briefly describe the comprehensive data available for this stand-alone currency system, as well as the most relevant background for interpreting our novel measures of money velocity for Sarafu across geographic locales and over time. 

\subsection{Data on transactions and account holders} \label{sec:data:data}

Grassroots Economics has made a portion of the 2020-21 system's administrative records available for research; the published dataset includes anonymized account information for around 55,000 users and records of all Sarafu transactions conducted from 25 January 2020 to 15 June 2021~\parencite{ruddick_sarafu_2021}. The 2020-21 Sarafu dataset can be accessed via the UK Data Service and is thoroughly documented~\parencite{mattsson_sarafu_2022}. The 2020-21 Sarafu dataset includes transactions totaling around 300 million Sarafu. 

Most user activity was recorded in the data as \textsc{standard} transactions. These would capture various economic and financial activities such as local purchases and participation in savings and lending groups.  We also consider \textsc{agent\_out} transactions to be ``transfers'' in that they indicate purchases, of sorts, that facilitated donations to savings and lending groups; there were limited instances of this. Generally, the exchange of Sarafu with Kenyan Shillings was not facilitated~\parencite{mattsson_sarafu_2022}.

The creation and removal of Sarafu took place via \textsc{disbursement} and \textsc{reclamation} transactions, respectively. New Sarafu was systematically issued to newly created accounts. Later in the observation period, so-called demurrage charges served to remove a small fraction of existing Sarafu each month. Besides routine currency management, various administrative operations were undertaken at times. The accounts involved in currency management and administrative operations are identified in the data with a \textit{system} label. We include in this category also the account labeled \textit{vendor}, since it was used to facilitate pandemic aid. Values reported in our figures and tables do not include the contribution of \textit{system} accounts. 

The 2020-21 Sarafu dataset also includes anonymized account data with several contextual attributes that describe the characteristics of account holders. The ``area name'' and ``business type'' are user-generated entries generalized into broader categories by staff at Grassroots Economics. They reflect the home locality of the user and the product category of the goods or services they provide to the community. Localities are also categorized into \textit{urban}, \textit{periurban}, and \textit{rural} ``area types.'' \textcite{mattsson_sarafu_2022} provide precise descriptions of these data fields and \textcite{ussher_complementary_2021} provide further detail on the values. 

\subsection{Information on shocks and interventions} \label{sec:data:info}

\begin{figure}[ht]
\centering
\includegraphics[width=0.46\linewidth]{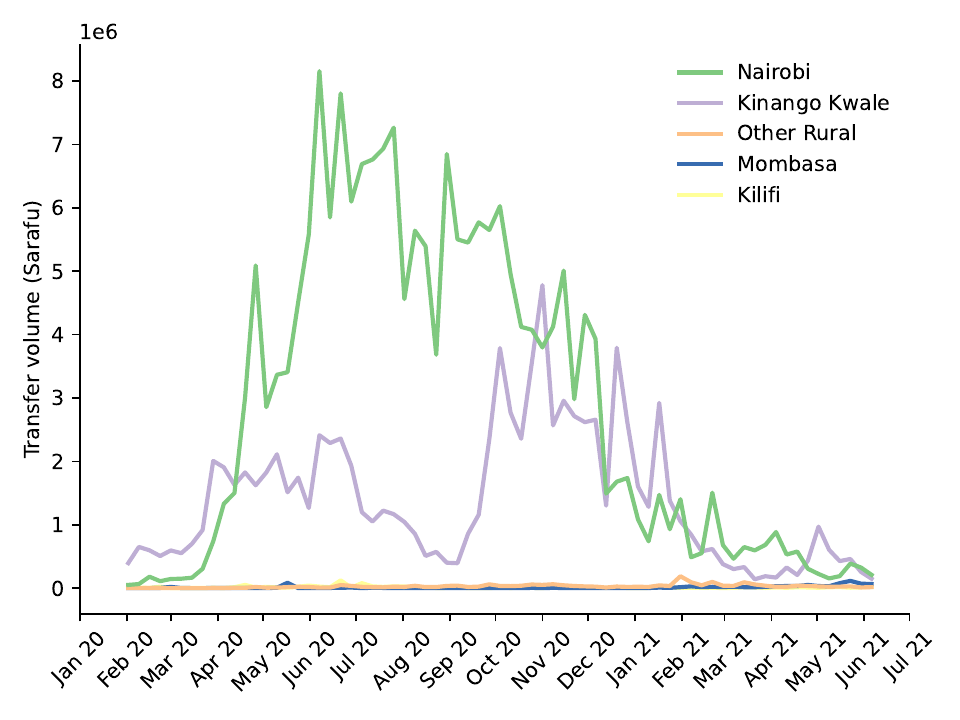}
\includegraphics[width=0.46\linewidth]{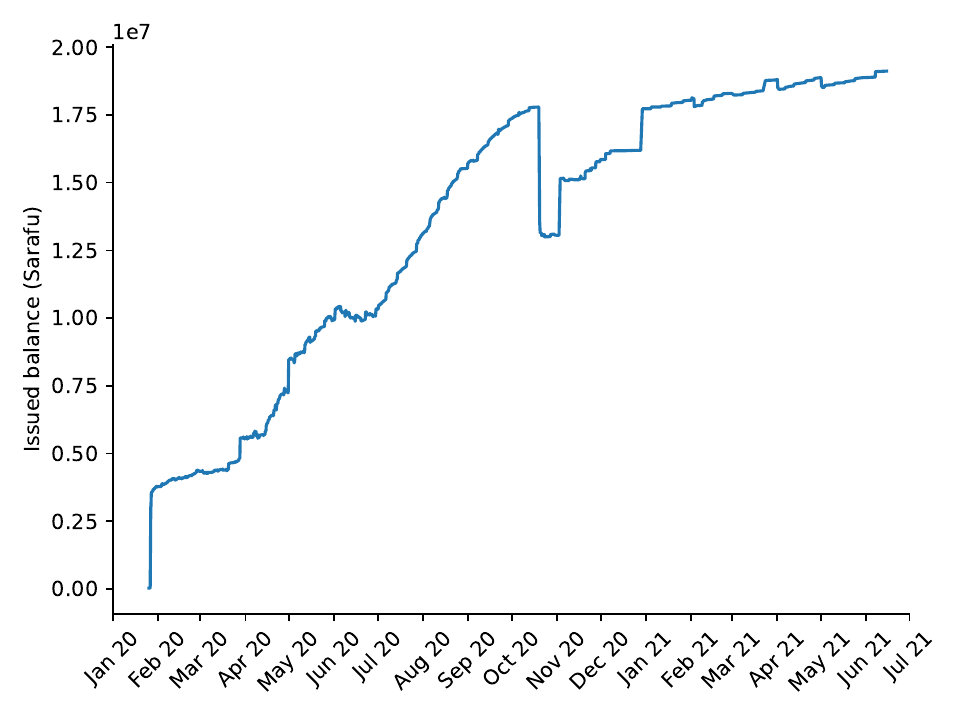}
\caption{Weekly transfer volumes per geographic area (left) and total issued balance (right) of Sarafu.}
\label{fig:sarafu}
\end{figure}

The 2020-21 Sarafu dataset covers the first year of the COVID-19 pandemic and the observation window includes several documented pilot projects, interventions, and currency operations~\parencite{ussher_complementary_2021,mattsson_sarafu_2022}. Figure~\ref{fig:sarafu} shows timeseries of Sarafu transfer volumes (left) and the total system balance (right). Transfer volumes expanded dramatically as the communities using Sarafu experienced economic disruption related to the COVID-19 pandemic. This pattern is in line with the prevailing understanding of community currencies, the use of which is counter-cyclical~\parencite{stodder_complementary_2009,stodder_macro-stability_2016,zeller_economic_2020}. The total issued balance of the Sarafu system also grew substantially in the first half of the observation period as new accounts were created. There are noticeable discontinuities where specific currency operations added or removed an appreciable fraction of the total balance.

Prior work with the 2020-21 Sarafu dataset has established that Sarafu was acting as a community currency in several areas of Kenya, seeing mostly local circulation and considerable involvement from community-based institutions. The structure of circulation within the Sarafu system over this period was highly modular and geographically localized, yet occurring among users providing diverse products~\parencite{mattsson_circulation_2023}. The explicit involvement of community-based institutions, especially savings and lending groups, is an innovative aspect of the Sarafu system. There are a few hundred \textit{group accounts} noted in the data, and they play an important role~\parencite{ba_cooperative_2023}. Moreover, digital community currencies are seen as a promising new modality for humanitarian aid. \textcite{ussher_complementary_2021} use the example of Sarafu to argue that digital community currencies compare favorably to cash assistance because they help establish local economic connections that keep money circulating within a community.

\section{Empirical analyses} \label{sec:results}

Measuring the average holding period of money opens up new avenues for empirical research. We illustrate three novel analyses made possible by our measure and measurement methodology, each on the Sarafu system. Section~\ref{sec:results:balance} considers the rate of circulation within the Sarafu system as a whole. We can compare the aggregate transfer velocity to the inverse of the average holding period and show the distribution of completed holding periods. Section~\ref{sec:results:disaggregation} demonstrates disaggregation in the context of localized pandemic-related shocks. We track the circulation of Sarafu over time, considering separately the communities using the currency in urban Nairobi and in rural Kinango Kwale. Section~\ref{sec:results:disaggregation} illustrates how measuring the average holding period offers a way to monitor money velocity while currency interventions are actively taking place. 

\subsection{Effective balance under non-stationary conditions} \label{sec:results:balance}

Measured in the conventional manner, unit tokens of Sarafu changed hands on average 0.31 times per week over the full observation period. However, the average holding period of Sarafu implies faster circulation than does the aggregate transfer velocity. Transfers were made with Sarafu that had been held for on average 1.36 weeks, corresponding to a non-stationary transfer velocity of 0.74 transactions per week. This discrepancy indicates that a substantial portion of the Sarafu balance was effectively static, sitting in accounts that spend at rates too low to contribute meaningfully towards the transfer volume. To produce the observed volume of transfers, effectively 42\% of the total Sarafu balance was circulating at 0.74 transactions per week.

\begin{figure}[htb]
\centering
\includegraphics[width=0.46\linewidth]{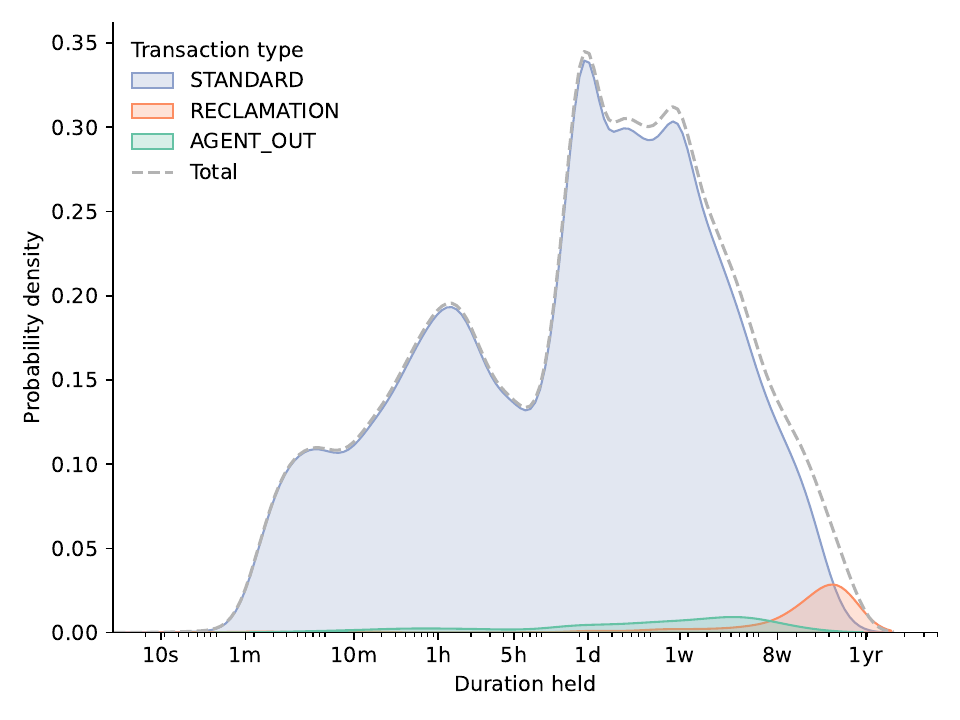}
\caption{Distribution of completed holding periods. Durations are shown for holding periods ending with Sarafu transactions observed between 25 January 2020 and 15 June 2021, normalized by the total transaction volume. Contributing to the total distribution (dashed line) are transactions of different types.}
\label{fig:distribution}
\end{figure}

Completed holding periods vary in duration over several orders of magnitude. Figure~\ref{fig:distribution} shows the probability density of observing periods with particular durations prior to a recorded transaction. Sarafu is often spent the same day it is received, sometimes in a matter of minutes. It is also common for token units to be held overnight and spent later in the same week, or in the following week. At times, Sarafu is held for longer periods before being spent. Recall from Section~\ref{sec:data:data} that we consider only \textsc{standard} and \textsc{agent\_out} transactions to contribute to the transfer volume. The average holding period as observed prior to transfer transactions is shorter than would be expected from the aggregate transfer velocity, leading a lower effective balance. Indeed, many of the longest-duration holding periods were observed prior to \textsc{reclamation} transactions. These were used to remove Sarafu from circulation. The long-duration tail of the total distribution thus reflects primarily currency management operations, not transfers initiated by spenders.

\subsection{Disaggregation under non-stationary conditions} \label{sec:results:disaggregation}

Changing circumstances can be expected to change spending patterns. In synthesizing much of the literature on heterogeneity in unemployment, inflation, and economic growth, \textcite{goodman-bacon_myth_2021} argues that without understanding the ways that different groups and individuals experience the economy, macroeconomic policy-makers cannot realize the full potential for economic growth. Measuring average holding periods allows us to analyze money velocity across disaggregated groups under non-stationary condition. Here we focus on geographic disaggregation, as Sarafu is a community currency with localized circulation~\parencite{mattsson_circulation_2023}. In the data set, accounts are labeled with a home locality as noted in Section~{\ref{sec:data}}.

Beginning in March 2020, widespread behavioral change and national mitigation policies contra the COVID-19 pandemic strongly affected transport, mobility, and business operations in Kenya. This resulted in substantial economic disruption, particularly in poor urban areas~\parencite{fews_net_kenya_2020}. As noted in Section~\ref{sec:data:info}, the Sarafu system expanded as the COVID-19 pandemic arrived in Kenya. The total system balance grew as Sarafu was issued to new accounts. Table~\ref{tab:areas} summarizes the circulation of Sarafu within the communities using Sarafu in Nairobi, Mombasa, Kilifi, Kinango Kwale, and other rural areas using our measures computed over the full observation period. Total transfer volumes were especially large in Nairobi and in Kinango Kwale. Nairobi is the capital city of Kenya, where Sarafu was just being introduced~\parencite{mattsson_sarafu_2022}. Kinango is an administrative division in rural Kwale county where Grassroots Economics has had a presence for many years~\parencite{marion_voucher_2018}. In the next sections we consider closely the experience of Sarafu users in informal settlements of Nairobi and in Kinango Kwale 

\begin{table}[h]
\centering
\begin{tabular}{l|rrrrr}
Area            & Transfer volume & Holding period & Transfer velocity    & Effective balance & Issued balance \\
\hline
                & $F_T$, Sarafu  & $\bar{\uptau}_T$, weeks & $V_T$, per week        & $M_T$, Sarafu    & $M_{\text{avg}}$, Sarafu \\
Nairobi & 200.05m & 0.66 & 1.53 & 1.81m & 5.42m \\
Kinango Kwale & 98.49m & 2.74 & 0.36 & 3.73m & 7.34m \\
Other Rural & 2.04m & 2.25 & 0.44 & 0.06m & 0.28m \\
Kilifi & 0.96m & 2.57 & 0.39 & 0.03m & 0.17m \\
Mombasa & 0.92m & 3.54 & 0.28 & 0.04m & 0.33m \\
\hdashline
Total & 302.47m & 1.36 & 0.74 & 5.94m & 13.55m \\
\end{tabular}
\caption{Non-stationary measures of the circulation of Sarafu, for different areas of Kenya.  Reported as the issued balance in each area is the time-average of this value measured hourly.}
\label{tab:areas}
\end{table}

\subsubsection{Income shock in Urban Nairobi}

The COVID-19 pandemic impacted urban areas of Kenya directly and prompted a series of official mitigation policies: remote-work directives and nationwide school closures were announced on 15 March 2020; targeted restrictions took effect over the following week and a stringent nationwide curfew was imposed on 27 March; limitations on free movement into and out of Kenya's major cities began on 6 April and remained in effect until 7 July; the nationwide curfew was progressively relaxed on 7 June and on 29 September; bars were allowed to re-open on 29 September; schools were reopened for three grade-levels on 12 October 2020, and fully reopened on 4 January 2021 (Data from \url{www.health.go.ke/press-releases}). In acting early, Kenya averted a potentially devastating initial wave of infection. However, the economic impact was substantial and the Kenyan economy shrank in 2020, as measured by Gross Domestic Product (GDP).

The urban poor experienced especially severe economic disruption. Reduced movement and limitations on business operations led to widespread losses in employment and income-generating opportunities for poor urban households. At the same time, mobility restrictions and delays in cross-border trade raised prices for consumer goods including staple foods~\parencite{fews_net_kenya_2020}. In phone surveys conducted on 14 April 2020 in informal settlements of Nairobi, 81\% of respondents reported complete or partial loss of income (36\% and 45\%, respectively). 87\% of respondents reported increased household expenditures, especially on food~\parencite{noauthor_kenya_2020,pinchoff_gendered_2021}. By June, informal settlements of Nairobi were noted as an \emph{Area of Concern} in the Food Security Outlook for Kenya published by the Famine Early Warning Systems Network~\parencite{fews_net_kenya_2020}. Food insecurity persisted through August for many poor urban households and lingered into December for some households, even as the wider Kenyan economy had begun to recover~\parencite{fews_net_kenya_2020-1,fews_net_kenya_2020-2}.

\begin{figure}[htb]
\centering
\includegraphics[width=0.46\linewidth]{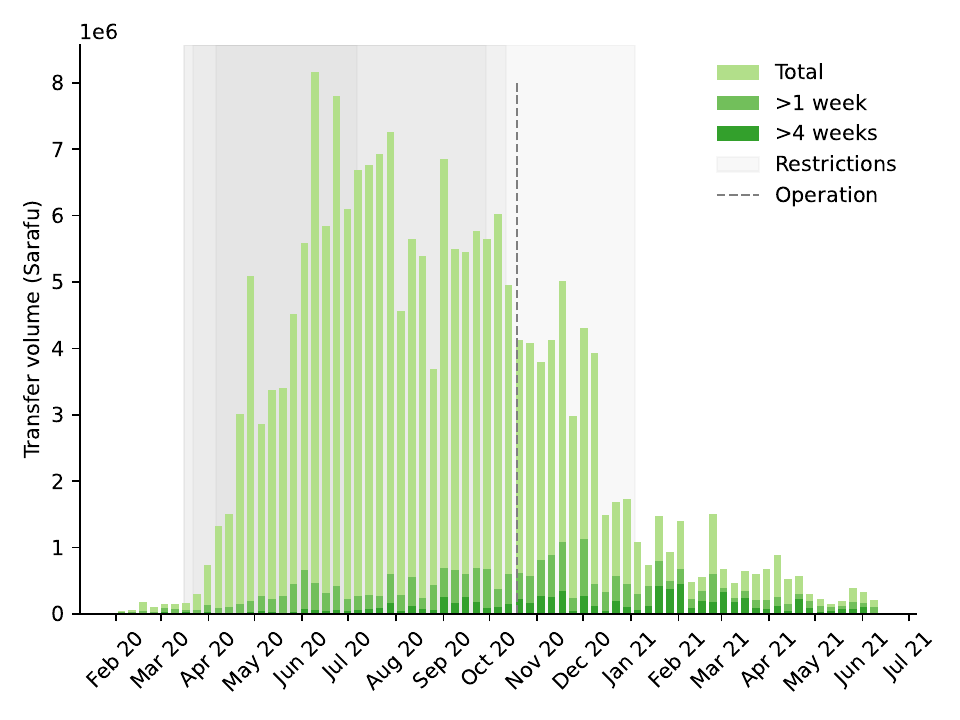}
\includegraphics[width=0.46\linewidth]{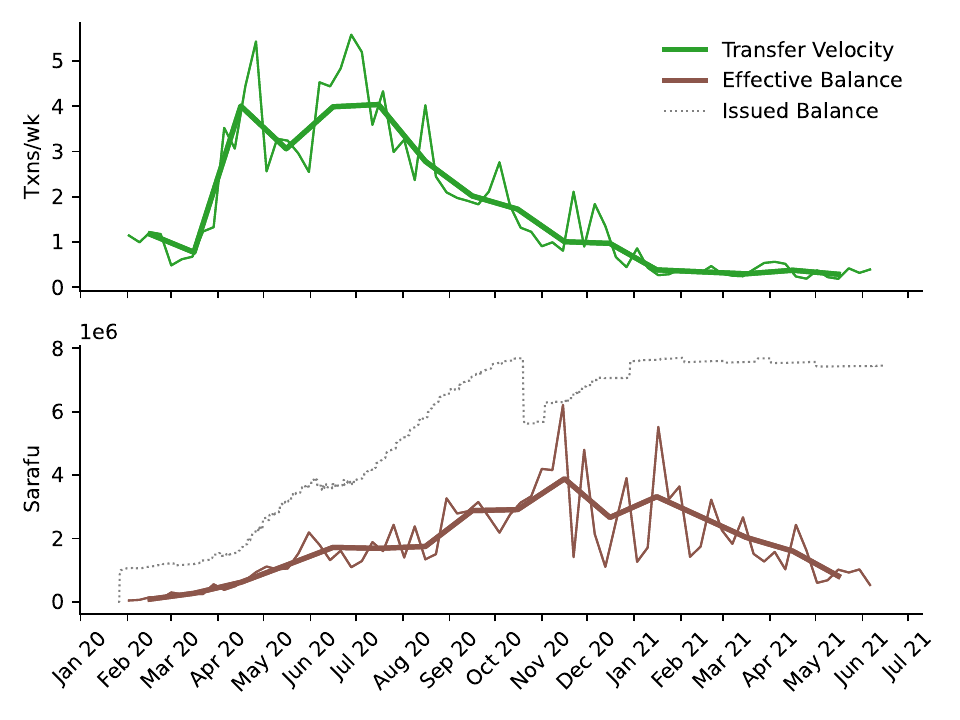}
\caption{The volume of transfers made each week by users registered in Nairobi (left) and our novel monetary indicators, weekly and monthly (right). The share of transfer volumes plotted in darker green are attributable to funds held for at least one or four weeks immediately prior to being transferred. Shaded in grey are periods with more restrictive COVID-19 pandemic mitigation policies. The grey vertical line gives the timing of a currency operation that removed a substantial share of Sarafu from the system.}
\label{fig:Nairobi}
\end{figure}

In one informal settlement of Nairobi, a village in the Mukuru slum, the initial disruption coincided with a targeted introduction of the Sarafu community currency~\parencite{mattsson_sarafu_2022}. Promotion, education, and training programs managed by the Kenyan Red Cross began in April 2020. Sarafu became part of an improvised humanitarian response effort, apparently meeting an acute need in this community. Use of the Sarafu system expanded dramatically. Figure~\ref{fig:Nairobi} (left) shows the weekly transfer volume by accounts registered in Nairobi, with periods of more severe restrictions noted in darker shades of grey. Transfer volumes rose by several orders of magnitude in the weeks following the introduction of Sarafu in April and reached a peak in July as restrictions were eased. The use of Sarafu in Nairobi decreased towards the end of 2020, presumably as economic conditions normalized.

Our measures reveal that transfer volumes in Nairobi arose out of intense use of a small subset of the total Sarafu available to this community. Figure~\ref{fig:Nairobi} (right) shows the weekly and monthly measures of money velocity and effective balance for Nairobi. The velocity rose from below one to above three transactions per week in April 2020 and remained exceptionally high into July 2020. Indeed, we can see from Figure~\ref{fig:Nairobi} (left) that most of the transfer volume in April 2020 and over the subsequent months is attributable to funds held for less than a week before being re-transacted. The effective balance also grew substantially over this period, but it lagged the growth in the issued balance of Sarafu in Nairobi. Recall from Section~\ref{sec:data:data} that new Sarafu was created primarily via new accounts under the currency management procedures active in this period. The newly issued Sarafu did not immediately reach those most keen on spending Sarafu. 

The frenzied circulation of Sarafu in Nairobi began to slow around the time the most restrictive COVID-19 pandemic mitigation policies were lifted, in July 2020. However, the effective balance continued to grow and transfer volumes remained elevated through November 2020. We consider this an indication that the use of Sarafu broadened over this period. The effective balance of Sarafu circulating in Nairobi began to decline in January 2021, as schools opened fully and the economic recovery reached also the poorest urban households~\parencite{fews_net_kenya_2020-1,fews_net_kenya_2020-2}. Transfer volumes subsided and the transfer velocity, for the smaller share of Sarafu still in use, settled at around one transfer every three to four weeks.

\subsubsection{Expansion in Rural Kinango Kwale}

Kinango Kwale is a poor rural area where many experience seasonal food insecurity. Marginal agricultural enterprises provide seasonal work for laborers and own agricultural production provides seasonal support for subsistence, both subject to climate variability~\parencite{fews_net_kenya_2020,moalf_climate_2016}. It is common for residents to leave for nearby urban areas in search of income-generating opportunities~\parencite{marion_voucher_2018}. 

The impact of the COVID-19 pandemic and nationwide mitigation policies was less direct in Kinango Kwale. Rural livelihoods were indirectly affected via disruptions in the livelihoods of migrant workers in urban areas and changes in the prices of staple foods. Staff at Grassroots Economics liken the economic shock affecting Kinango Kwale to that seen during holiday periods: an influx of migrant workers returning home. In meeting the needs of additional people at higher prices, communities with existing access to the Sarafu system had the option to use the community currency and, as a result, the transfer velocity and the effective balance of Sarafu in Kinango Kwale show considerable temporal heterogeneity.

\begin{figure}[htb]
\centering
\includegraphics[width=0.46\linewidth]{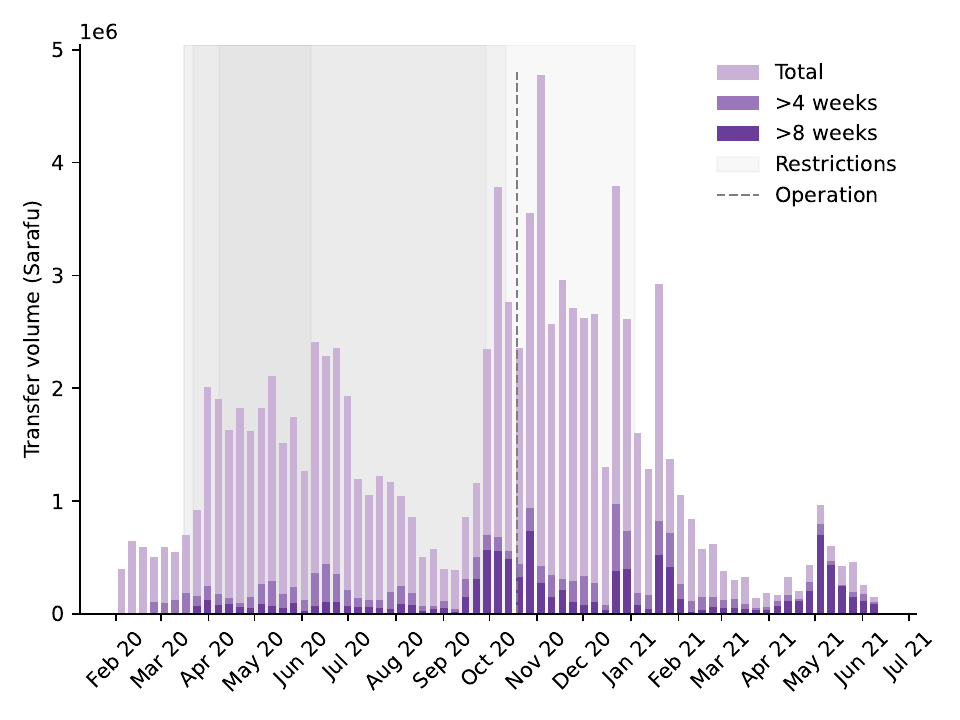}
\includegraphics[width=0.46\linewidth]{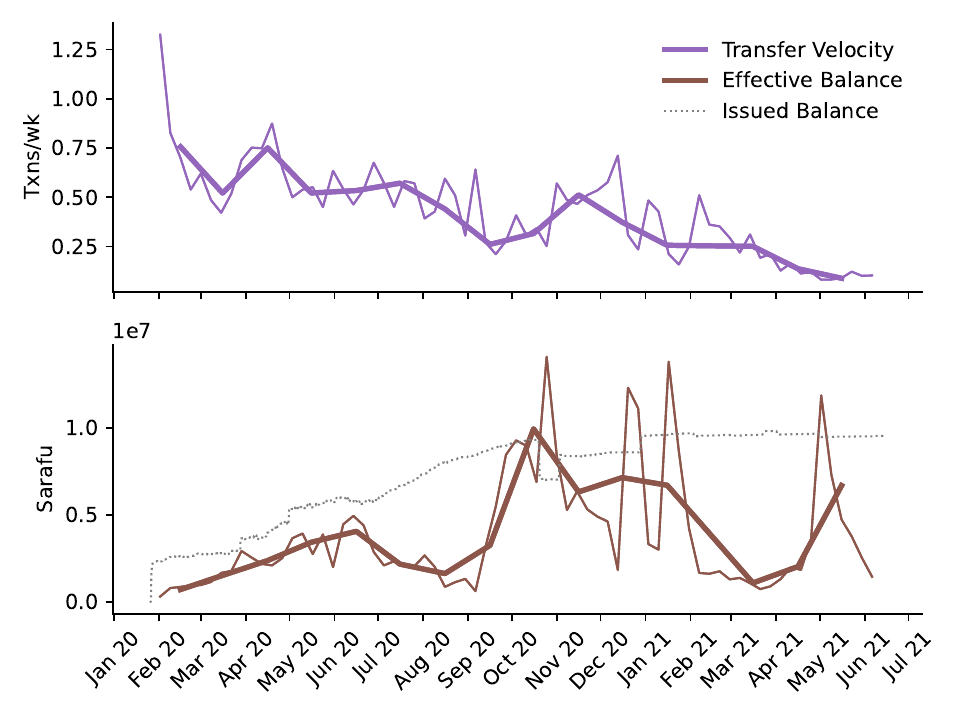}
\caption{The volume of transfers made each week by users registered in Kinango Kwale (left) and our novel monetary indicators, weekly and monthly (right). Transfer volumes plotted in darker shades are attributable to funds held for more than four or eight weeks immediately prior to being transferred. Shaded in grey are periods with more restrictive COVID-19 pandemic mitigation policies. The grey vertical line gives the timing of a currency operation that removed a substantial share of Sarafu from the system.}
\label{fig:KinangoKwale}
\end{figure}

Figure~\ref{fig:KinangoKwale} (left) shows the volume of transfers made each week by users registered in Kinango Kwale and Figure~\ref{fig:KinangoKwale} (right) presents the inverse of the average holding period and the effective balance, weekly and monthly. Sarafu saw a sharp increase in use in Kinango Kwale beginning in late March 2020---weekly transfer volumes tripled between February and April. This coincides with a temporary spike in the money velocity, suggesting that already-active users stepped up their use of Sarafu as nationwide restrictions entered into effect. High transfer volumes were then sustained for several months when the effective balance increased. Perhaps because of existing familiarity with Sarafu in the community, large amounts of Sarafu issued to new users reached those keen on using in. 

Limitations on free movement were lifted for Kwale county on 7 June, though they remained in effect in Kenya's major cities until 7 July. These developments and, especially, a good harvest in July and August lessened the economic stressors affecting marginal agricultural areas of eastern Kenya~\parencite{fews_net_kenya_2020-1}. Use of Sarafu in Kinango Kwale reached a lull in September, with average circulation rate and the effective balance falling in tandem. This is a similar pattern as that which occurred in Nairobi when economic conditions normalized, occurring some months earlier.

Intriguingly, Kinango Kwale saw a second period of increased use beginning in late September 2020. This swell in transfer volume is of a decidedly different character: we initially see a spike not in the Sarafu velocity but in the effective balance, meaning that rising transfer volumes initially reflected renewed use by {\it less} active accounts. Higher transfer volumes were sustained for over the following months in that the money velocity subsequently rose. There are two explanations for this resurgence. It may simply reflect local seasonality in livelihoods and heightened food insecurity~\parencite{fews_net_kenya_2020-1,fews_net_kenya_2020-2} consistent with the prevailing understanding that community currencies see more use in difficult times~\parencite{stodder_complementary_2009,zeller_economic_2020}. In the following section we consider also a second explanation, wherein communities in Kinango Kwale may have been responding logically to a specific intervention by Grassroots Economics.

\subsection{Measurement during interventions} \label{sec:results:intervention}

The average holding period of money can be measured even as interventions take place within a currency system. On 19 October 2020, Grassroots Economics initiated a large set of \textsc{reclamation} transactions that removed Sarafu from a large set of accounts~\parencite{mattsson_sarafu_2022}. This operation dissolved 26.7\% of all Sarafu token units over two days, visible as a near-instantaneous drop in the total issued balance in Figure~\ref{fig:sarafu} (right). The conventional measure of the aggregate transfer velocity of Sarafu would be affected, by definition, via the time-average of the total issued balance as in Eqn.~\ref{eqn:velocity_conv}. The average holding period is not necessarily affected. Completed holding periods are observed when users make transactions and their average duration is affected only to the extent the intervention impacts user activity. 

The October 2020 intervention was targeted primarily at inactive accounts and, secondarily, at accounts with large balances. This currency operation accounted for 66.2\% of all Sarafu removed via \textsc{reclamation} transactions over the entire observation window. It removed predominantly long-held funds, as can be seen in Figure~\ref{fig:distribution}. These units of Sarafu had been sitting in the same account for, on average, 25.1 weeks prior to being dissolved; 89.8\% had never been transferred. \textcite{mattsson_sarafu_2022} report that the intention to close out inactive accounts was announced by Grassroots Economics ahead of the operation itself.

Our measures reveal the absence of a strong response by spenders in Nairobi to the announcement and implementation of the October 2020 currency operation. A total of 2.1 million Sarafu was removed from accounts in Nairobi, corresponding to 27\% of their total issued balance. Neither the average circulation rate nor the effective balance deviated discernibly from their trends in September/October 2020 (Figure~\ref{fig:Nairobi}, right). The transfer velocity was falling and the effective balance was rising, both gradually. The transfer volume remained large in Nairobi through September and October 2020; the share of transfers made with long-held funds remained small (Figure~\ref{fig:Nairobi}, left). We see no discernible response from Sarafu users in Nairobi.

In Kinango Kwale, communities using Sarafu appear to have responded more strongly. A total of 2.2 million Sarafu was removed from accounts in Kinango Kwale, corresponding to 24\% of their total issued balance. We see from Figure~\ref{fig:KinangoKwale} that transfers with older funds increased in late September and early October 2020, a pattern that is consistent with a robust response to the announcement of a future intervention targeting inactive accounts. Unfortunately, the timing of the announcement is not precisely known. The same pattern might also arise from renewed demand for Sarafu in a period of seasonal food insecurity. In any case, older Sarafu re-entering active circulation is captured by our measures as a jump in the effective balance. Indeed, the effective balance in Kinango Kwale (9.9 million Sarafu) exceeded the issued balance in Kinango Kwale (8.4 million Sarafu) in the month of October 2020. This is remarkably different from the pattern observed in this area five or six months earlier, where transfer volumes picked up due to a jump in the transfer velocity. Measuring the average holding period of Sarafu lets us tell these patterns apart.

\section{Conclusion} \label{sec:conclusion} 

Our work has demonstrated how the rate at which money changes hands can be measured as the inverse of its average holding period. This generalizes the conventional definition of the transfer velocity of money to non-stationary conditions. Our measurement methodology computes this value directly from the empirical distribution of completed holding periods, which can be extracted from the transaction data recorded by payment systems themselves. This enables empirical analysis of money velocity under disaggregation. Our empirical analyses focused on Sarafu, a self-contained example where dynamics of use, external shocks, and interventions give rise to non-stationary conditions. We find that the circulation of Sarafu was deeply heterogeneous in ways that are unlikely to be unique to Sarafu. Moreover, we show how our measure gives meaningful insights during shocks and while policy interventions are taking place.   

Our methodological advance contributes to a growing literature that acknowledges heterogeneity in how groups and individuals experience the economy and seeks out novel data that allows for disaggregated analysis. For monetary analysis, transaction records open up a new frontier. Our approach can be used in the future to conduct analyses on other digital payment systems, such as those operated by mobile payment providers, commercial banks, and central banks. In dispensing with the assumption that these real-world systems operate under stationary conditions we make it possible to consider money velocity, empirically, even when payment processing is highly fragmented or rapidly changing. Finally, our generalized monetary indicators remain properly defined when the monetary base expands or contracts suddenly. This can lead to improved monitoring of the money supply to better inform macroeconomic policy.

%% Back matter %%

%\section*{Conflict of Interest Statement}

%\section*{Author Contributions}

%\section*{Funding}

\section*{Acknowledgments}

The authors would like to thank several anonymous referees for their constructive comments and suggestions, which have significantly improved the work. We are grateful to William Ruddick and others at Grassroots Economics Foundation for sharing their knowledge with us and their records with the research community. In addition, we thank Carlo Campajola, Alastair Langtry, the participants of the Cambridge Janeway Networks Webinar Series, researchers at the Bank Indonesia Institute, the participants of the 2022 Workshop on Data-Driven Economic Agent-Based Models at Sant'Anna School of Advanced Studies in Pisa, as well as participants of NetSci 2023 in Vienna and CCS 2023 in Salvador for their helpful discussions and comments.

\section*{Data Availability Statement}

The Sarafu CIC 2020–2021 dataset, is available via the UK Data Service (UKDS) ReShare platform~\parencite{ruddick_sarafu_2021}. Note that the dataset is “safeguarded”, meaning that access is limited to those who have registered with the UKDS. The supplementary material includes instructions on applying for an account and downloading the data. Re-use is constrained by the UKDS End User License (EUL); re-identifying individuals is prohibited. 

\section*{Supplementary Material}

The code and configuration files required to reproduce our analysis is available at \url{https://github.com/carolinamattsson/transfer-velocity-of-money}. 

%% References %%

%\bibliographystyle{frontiersinSCNS_ENG_HUMS}
%\bibliography{references}
\printbibliography

@article{leontief_money-flow_1993,
	title = {Money-flow Computations},
	volume = {5},
	issn = {0953-5314, 1469-5758},
	url = {http://www.tandfonline.com/doi/full/10.1080/09535319300000019},
	doi = {10.1080/09535319300000019},
	pages = {225--233},
	number = {3},
	journaltitle = {Economic Systems Research},
	shortjournal = {Economic Systems Research},
	author = {Leontief, Wassily and Brody, András},
	urldate = {2019-06-18},
	date = {1993-01},
	langid = {english},
}

@article{wang_prospects_2005,
	title = {Prospects for Money Transfer Models},
	url = {http://arxiv.org/abs/physics/0507161},
	journaltitle = {{arXiv}:physics/0507161},
	author = {Wang, Yougui and Ding, Ning and Xi, Ning},
	date = {2005-07-21},
	eprinttype = {arxiv},
	eprint = {physics/0507161},
}

@article{wang_circulation_2003,
	title = {The circulation of money and holding time distribution},
	volume = {324},
	issn = {0378-4371},
	url = {http://www.sciencedirect.com/science/article/pii/S0378437103000748},
	doi = {10.1016/S0378-4371(03)00074-8},
	pages = {665--677},
	number = {3},
	journaltitle = {Physica A: Statistical Mechanics and its Applications},
	shortjournal = {Physica A: Statistical Mechanics and its Applications},
	author = {Wang, Yougui and Ding, Ning and Zhang, Li},
	date = {2003-06-15},
}

@article{kanazawa_kinetic_2018,
	title = {Kinetic theory for financial Brownian motion from microscopic dynamics},
	volume = {98},
	url = {https://link.aps.org/doi/10.1103/PhysRevE.98.052317},
	doi = {10.1103/PhysRevE.98.052317},
	pages = {052317},
	number = {5},
	journaltitle = {Physical Review E},
	shortjournal = {Phys. Rev. E},
	author = {Kanazawa, Kiyoshi and Sueshige, Takumi and Takayasu, Hideki and Takayasu, Misako},
	urldate = {2018-11-30},
	date = {2018-11-29},
}

@report{bartscher_monetary_2021,
	location = {New York, {NY}},
	title = {Monetary Policy and Racial Inequality},
	url = {https://www.minneapolisfed.org:443/research/institute-working-papers/monetary-policy-and-racial-inequality},
	number = {959},
	institution = {Federal Reserve Bank of New York},
	author = {Bartscher, Alina and Kuhn, Moritz and Schularick, Moritz and Wachtel, Paul},
	urldate = {2022-01-13},
	date = {2021-02-04},
}

@report{gornemann_doves_2021,
	location = {Germany},
	title = {Doves for the Rich, Hawks for the Poor? Distributional Consequences of Systematic Monetary Policy},
	url = {https://www.minneapolisfed.org:443/research/institute-working-papers/doves-for-the-rich-hawks-for-the-poor-distributional-consequences-of-systematic-monetary-policy},
	shorttitle = {Doves for the Rich, Hawks for the Poor?},
	number = {089},
	institution = {University of Bonn and University of Cologne},
	author = {Gornemann, Nils and Kuester, Keith and Nakajima, Makoto},
	urldate = {2022-01-13},
	date = {2021-06-29},
}

@article{stodder_macro-stability_2016,
	title = {The Macro-Stability of Swiss {WIR}-Bank Credits: Balance, Velocity, and Leverage},
	volume = {58},
	url = {https://ideas.repec.org/a/pal/compes/v58y2016i4d10.1057_s41294-016-0001-5.html},
	doi = {10.1057/s41294-016-0001-5},
	shorttitle = {The Macro-Stability of Swiss {WIR}-Bank Credits},
	pages = {570--605},
	number = {4},
	journaltitle = {Comparative Economic Studies},
	author = {Stodder, James and Lietaer, Bernard},
	urldate = {2021-03-16},
	date = {2016},
	langid = {english},
	note = {Publisher: Palgrave Macmillan \& Association for Comparative Economic Studies},
}

@article{campajola_microvelocity_2022,
	title = {{MicroVelocity}: rethinking the Velocity of Money for digital currencies},
	url = {http://arxiv.org/abs/2201.13416},
	shorttitle = {{MicroVelocity}},
	journaltitle = {{arXiv}:2201.13416 [physics, q-fin]},
	author = {Campajola, Carlo and D'Errico, Marco and Tessone, Claudio J.},
	urldate = {2022-03-01},
	date = {2022-01-31},
	eprinttype = {arxiv},
	eprint = {2201.13416},
}

@article{goodman-bacon_myth_2021,
	title = {The myth of the rising tide {\textbar} Federal Reserve Bank of Minneapolis},
	url = {https://www.minneapolisfed.org:443/article/2021/the-myth-of-the-rising-tide},
	author = {Goodman-Bacon, Andrew},
	urldate = {2022-01-13},
	date = {2021-04-14},
}

@article{argente_cost_2021,
	title = {Cost of Living Inequality During the Great Recession},
	volume = {19},
	issn = {1542-4766},
	url = {https://doi.org/10.1093/jeea/jvaa018},
	doi = {10.1093/jeea/jvaa018},
	pages = {913--952},
	number = {2},
	journaltitle = {Journal of the European Economic Association},
	shortjournal = {Journal of the European Economic Association},
	author = {Argente, David and Lee, Munseob},
	urldate = {2022-01-13},
	date = {2021-04-01},
}

@article{leao_why_2005,
	title = {Why does the velocity of money move pro‐cyclically?},
	volume = {19},
	issn = {0269-2171},
	url = {https://doi.org/10.1080/0269217042000312641},
	doi = {10.1080/0269217042000312641},
	pages = {119--135},
	number = {1},
	journaltitle = {International Review of Applied Economics},
	author = {Leão, Pedro},
	urldate = {2022-01-12},
	date = {2005-01-01},
	note = {Publisher: Routledge
\_eprint: https://doi.org/10.1080/0269217042000312641},
}

@online{federal_reserve_bank_of_st_louis_velocity_2022,
	title = {Velocity of M2 Money Stock},
	url = {https://fred.stlouisfed.org/series/M2V},
	shorttitle = {M2V},
	titleaddon = {{FRED}, Federal Reserve Bank of St. Louis},
	author = {{Federal Reserve Bank of St. Louis}},
	urldate = {2022-01-12},
	date = {2022-01-03},
	note = {Publisher: {FRED}, Federal Reserve Bank of St. Louis},
}

@article{mbiti_home_2013,
	title = {The Home Economics of E-Money: Velocity, Cash Management, and Discount Rates of M-Pesa Users},
	volume = {103},
	issn = {0002-8282},
	url = {http://pubs.aeaweb.org/doi/10.1257/aer.103.3.369},
	doi = {10.1257/aer.103.3.369},
	shorttitle = {The Home Economics of E-Money},
	pages = {369--374},
	number = {3},
	journaltitle = {American Economic Review},
	author = {Mbiti, Isaac and Weil, David N},
	urldate = {2018-10-09},
	date = {2013-05},
	langid = {english},
}

@incollection{mbiti_mobile_2015,
	title = {Mobile banking: The impact of M-Pesa in Kenya},
	shorttitle = {Mobile banking},
	pages = {247--293},
	booktitle = {African Successes, Volume {III}: Modernization and Development},
	publisher = {University of Chicago Press},
	author = {Mbiti, Isaac and Weil, David N.},
	date = {2015},
}

@online{ruddick_sarafu_2021,
	title = {Sarafu Community Inclusion Currency, 2020-2021},
	rights = {ukda\_eul},
	url = {https://reshare.ukdataservice.ac.uk/855142/},
	type = {Data Collection},
	author = {Ruddick, William O.},
	editora = {Criscione, Teodoro and Mattsson, C. E. S.},
	editoratype = {collaborator},
	urldate = {2023-03-22},
	date = {2021-08-31},
	langid = {english},
	doi = {10.5255/UKDA-SN-855142},
	note = {Publisher: {UK} Data Service},
}

@report{noauthor_kenya_2020,
	location = {Nairobi},
	title = {Kenya: {COVID}-19 knowledge, attitudes, practices and needs—Responses from second round of data collection in five Nairobi informal settlements (Kibera, Huruma, Kariobangi, Dandora, Mathare)},
	url = {https://knowledgecommons.popcouncil.org/departments_sbsr-pgy/974},
	shorttitle = {Kenya},
	institution = {Population Council},
	type = {{COVID}-19 Research \& Evaluations Brief},
	date = {2020-04-22},
}

@report{fews_net_kenya_2020,
	location = {Kenya},
	title = {Kenya Food Security Outlook Update, August 2020},
	url = {https://reliefweb.int/report/kenya/kenya-food-security-outlook-update-august-2020},
	institution = {Famine Early Warning Systems Network},
	author = {{FEWS NET}},
	urldate = {2023-05-04},
	date = {2020-09-08},
	langid = {english},
}

@article{mattsson_circulation_2023,
	title = {Circulation of a digital community currency},
	volume = {13},
	rights = {2023 The Author(s)},
	issn = {2045-2322},
	url = {https://www.nature.com/articles/s41598-023-33184-1},
	doi = {10.1038/s41598-023-33184-1},
	pages = {Article number: 5864},
	journaltitle = {Scientific Reports},
	shortjournal = {Sci Rep},
	author = {Mattsson, Carolina E. S. and Criscione, Teodoro and Takes, Frank W.},
	urldate = {2023-04-13},
	date = {2023-04-11},
	langid = {english},
	note = {Article number: 5864},
}

@report{fews_net_kenya_2020-1,
	location = {Kenya},
	title = {Kenya Food Security Outlook Update, December 2020},
	url = {https://reliefweb.int/report/kenya/kenya-food-security-outlook-update-december-2020},
	institution = {Famine Early Warning Systems Network},
	author = {{FEWS NET}},
	urldate = {2023-05-12},
	date = {2020-12-31},
	langid = {english},
}

@report{fews_net_kenya_2020-2,
	location = {Kenya},
	title = {Kenya Food Security Outlook, June 2020 to January 2021},
	url = {https://reliefweb.int/report/kenya/kenya-food-security-outlook-june-2020-january-2021},
	institution = {Famine Early Warning Systems Network},
	author = {{FEWS NET}},
	urldate = {2023-05-12},
	date = {2020-07-11},
	langid = {english},
}

@article{pinchoff_gendered_2021,
	title = {Gendered economic, social and health effects of the {COVID}-19 pandemic and mitigation policies in Kenya: evidence from a prospective cohort survey in Nairobi informal settlements},
	volume = {11},
	rights = {© Author(s) (or their employer(s)) 2021. Re-use permitted under {CC} {BY}. Published by {BMJ}.. https://creativecommons.org/licenses/by/4.0/This is an open access article distributed in accordance with the Creative Commons Attribution 4.0 Unported ({CC} {BY} 4.0) license, which permits others to copy, redistribute, remix, transform and build upon this work for any purpose, provided the original work is properly cited, a link to the licence is given, and indication of whether changes were made. See: https://creativecommons.org/licenses/by/4.0/.},
	issn = {2044-6055, 2044-6055},
	url = {https://bmjopen.bmj.com/content/11/3/e042749},
	doi = {10.1136/bmjopen-2020-042749},
	shorttitle = {Gendered economic, social and health effects of the {COVID}-19 pandemic and mitigation policies in Kenya},
	pages = {e042749},
	number = {3},
	journaltitle = {{BMJ} Open},
	author = {Pinchoff, Jessie and Austrian, Karen and Rajshekhar, Nandita and Abuya, Timothy and Kangwana, Beth and Ochako, Rhoune and Tidwell, James Benjamin and Mwanga, Daniel and Muluve, Eva and Mbushi, Faith and Nzioki, Mercy and Ngo, Thoai D.},
	urldate = {2023-05-12},
	date = {2021-03-01},
	langid = {english},
	pmid = {33658260},
	note = {Publisher: British Medical Journal Publishing Group
Section: Global health},
}

@article{mattsson_sarafu_2022,
	title = {Sarafu Community Inclusion Currency 2020-2021},
	volume = {9},
	doi = {https://doi.org/10.1038/s41597-022-01539-4},
	number = {426},
	journaltitle = {Scientific Data},
	author = {Mattsson, Carolina E. S. and Criscione, Teodoro and Ruddick, William O.},
	date = {2022-07},
}

@thesis{marion_voucher_2018,
	title = {Voucher Systems for Food Security: A Case Study on Kenya's Sarafu-Credit},
	url = {https://www.researchgate.net/publication/323550475_Voucher_Systems_for_Food_Security_A_Case_Study_on_Kenya%27s_Sarafu-Credit_Working_paper},
	shorttitle = {Voucher Systems for Food Security},
	institution = {University of Copenhagen},
	type = {{MS} Thesis},
	author = {Marion, Cauvet},
	date = {2018-03-05},
	note = {10.13140/{RG}.2.2.26399.05282},
}

@article{benati_money_2020,
	title = {Money velocity and the natural rate of interest},
	volume = {116},
	issn = {0304-3932},
	url = {https://www.sciencedirect.com/science/article/pii/S0304393219301709},
	doi = {10.1016/j.jmoneco.2019.09.012},
	pages = {117--134},
	journaltitle = {Journal of Monetary Economics},
	shortjournal = {Journal of Monetary Economics},
	author = {Benati, Luca},
	urldate = {2023-07-26},
	date = {2020-12-01},
	langid = {english},
}

@article{benati_new_2023,
	title = {A {New} {Approach} to {Estimating} the {Natural} {Rate} of {Interest}},
	volume = {n/a},
	copyright = {© 2023 The Ohio State University.},
	issn = {1538-4616},
	url = {https://onlinelibrary.wiley.com/doi/abs/10.1111/jmcb.13013},
	doi = {10.1111/jmcb.13013},
	number = {n/a},
    date = {2023-01},
	urldate = {2023-08-11},
	journal = {Journal of Money, Credit and Banking},
	author = {Benati, Luca},
}

@incollection{aladangady_transaction_2022,
	title = {From {Transaction} {Data} to {Economic} {Statistics}: {Constructing} {Real}-{Time}, {High}-{Frequency}, {Geographic} {Measures} of {Consumer} {Spending}},
	isbn = {978-0-226-80139-1},
	shorttitle = {From {Transaction} {Data} to {Economic} {Statistics}},
	url = {https://www.nber.org/books-and-chapters/big-data-twenty-first-century-economic-statistics/transaction-data-economic-statistics-constructing-real-time-high-frequency-geographic-measures},
	urldate = {2025-03-17},
	booktitle = {Big {Data} for {Twenty}-{First}-{Century} {Economic} {Statistics}},
	publisher = {University of Chicago Press},
	author = {Aladangady, Aditya and Aron-Dine, Shifrah and Dunn, Wendy and Feiveson, Laura and Lengermann, Paul and Sahm, Claudia},
	month = feb,
	year = {2022},
	pages = {115--145},
}

@article{waskom_seaborn_2021,
    doi = {10.21105/joss.03021},
    url = {https://doi.org/10.21105/joss.03021},
    year = {2021},
    publisher = {The Open Journal},
    volume = {6},
    number = {60},
    pages = {3021},
    author = {Michael L. Waskom},
    title = {seaborn: statistical data visualization},
    journal = {Journal of Open Source Software}
 }

@article{stodder_complementary_2009,
	title = {Complementary credit networks and macroeconomic stability: {Switzerland}'s {Wirtschaftsring}},
	volume = {72},
	issn = {0167-2681},
	shorttitle = {Complementary credit networks and macroeconomic stability},
	url = {https://www.sciencedirect.com/science/article/pii/S0167268109001772},
	doi = {10.1016/j.jebo.2009.06.002},
	number = {1},
	urldate = {2021-08-26},
	journal = {Journal of Economic Behavior \& Organization},
	author = {Stodder, James},
	month = oct,
	year = {2009},
	pages = {79--95},
}

@article{zeller_economic_2020,
	title = {Economic {Advantages} of {Community} {Currencies}},
	volume = {13},
	copyright = {http://creativecommons.org/licenses/by/3.0/},
	url = {https://www.mdpi.com/1911-8074/13/11/271},
	doi = {10.3390/jrfm13110271},
	number = {11},
	urldate = {2021-08-26},
	journal = {Journal of Risk and Financial Management},
	author = {Zeller, Sarah},
	month = nov,
	year = {2020},
	not2e = {Number: 11
Publisher: Multidisciplinary Digital Publishing Institute},
	pages = {271},
}

@article{ussher_complementary_2021,
	title = {Complementary {Currencies} for {Humanitarian} {Aid}},
	volume = {14},
	copyright = {http://creativecommons.org/licenses/by/3.0/},
	url = {https://www.mdpi.com/1911-8074/14/11/557},
	doi = {10.3390/jrfm14110557},
	number = {11},
	urldate = {2021-11-18},
	journal = {Journal of Risk and Financial Management},
	author = {Ussher, Leanne and Ebert, Laura and Gómez, Georgina M. and Ruddick, William O.},
	month = nov,
	year = {2021},
	pages = {557},
}

@misc{reback_pandas-devpandas_2020,
	title = {pandas-dev/pandas: {Pandas} v0.24.2},
	url = {https://doi.org/10.5281/zenodo.3509134},
	publisher = {Zenodo},
	author = {Reback, Jeff and McKinney, Wes and {jbrockmendel} and Bossche, Joris Van den and Augspurger, Tom and Cloud, Phillip and {gfyoung} and {Sinhrks} and Klein, Adam and Roeschke, Matthew and Hawkins, Simon and Tratner, Jeff and She, Chang and Ayd, William and Petersen, Terji and Garcia, Marc and Schendel, Jeremy and Hayden, Andy and {MomIsBestFriend} and Jancauskas, Vytautas and Battiston, Pietro and Seabold, Skipper and {chris-b1} and {h-vetinari} and Hoyer, Stephan and Overmeire, Wouter and {alimcmaster1} and Dong, Kaiqi and Whelan, Christopher and Mehyar, Mortada},
	month = feb,
	year = {2020},
	doi = {10.5281/zenodo.3509134}
}

@misc{caswell_matplotlibmatplotlib_2019,
	title = {matplotlib/matplotlib v3.1.0},
	url = {https://doi.org/10.5281/zenodo.2893252},
	publisher = {Zenodo},
	author = {Caswell, Thomas A. and Droettboom, Michael and Hunter, John and Firing, Eric and Lee, Antony and Klymak, Jody and Stansby, David and Andrade, Elliott Sales de and Nielsen, Jens Hedegaard and Varoquaux, Nelle and Root, Benjamin and Hoffmann, Tim and Elson, Phil and May, Ryan and Dale, Darren and Lee, Jae-Joon and Seppänen, Jouni K. and McDougall, Damon and Straw, Andrew and Hobson, Paul and Gohlke, Christoph and Yu, Tony S. and Ma, Eric and Vincent, Adrien F. and Silvester, Steven and Moad, Charlie and Katins, Jan and Kniazev, Nikita and Ariza, Federico and Ernest, Elan},
	month = may,
	year = {2019},
	doi = {10.5281/zenodo.2893252}
}

@misc{mattsson_carolinamattssonfollow--money_2020,
	title = {carolinamattsson/follow-the-money v0.2.0},
	url = {https://github.com/carolinamattsson/follow-the-money},
	author = {Mattsson, Carolina},
	month = oct,
	year = {2020}
}

@article{mattsson_trajectories_2021,
	title = {Trajectories through temporal networks},
	volume = {6},
	copyright = {2021 The Author(s)},
	issn = {2364-8228},
	url = {https://appliednetsci.springeropen.com/articles/10.1007/s41109-021-00374-7},
	doi = {10.1007/s41109-021-00374-7},
	number = {1},
	urldate = {2021-06-02},
	journal = {Applied Network Science},
	author = {Mattsson, Carolina E. S. and Takes, Frank W.},
	month = dec,
	year = {2021},
	no2te = {Number: 1
Publisher: SpringerOpen},
	pages = {1--31},
}

@article{elbahrawy_evolutionary_2017,
	title = {Evolutionary dynamics of the cryptocurrency market},
	volume = {4},
	number = {11},
	journal = {Royal Society open science},
	author = {ElBahrawy, Abeer and Alessandretti, Laura and Kandler, Anne and Pastor-Satorras, Romualdo and Baronchelli, Andrea},
	year = {2017},
	no2te = {Publisher: The Royal Society Publishing},
	pages = {170623},
}

@techreport{moalf_climate_2016,
	address = {Nairobi, Kenya},
	title = {Climate {Risk} {Profile} for {Kwale} {County}},
	url = {https://cgspace.cgiar.org/bitstream/handle/10568/80456/Kwale_Climate\%20Risk\%20Profile.pdf},
	institution = {Ministry of Agriculture, Livestock and Fisheries of Kenya},
	author = {{MoALF}},
	year = {2016},
}

@article{ba_cooperative_2023,
	title = {Cooperative behavior in blockchain-based complementary currency networks through time: {The} {Sarafu} case study},
	shorttitle = {Cooperative behavior in blockchain-based complementary currency networks through time},
	journal = {Future Generation Computer Systems},
	author = {Ba, Cheick Tidiane and Zignani, Matteo and Gaito, Sabrina},
	year = {2023},
}

@article{friedman_theoretical_1970,
	title = {A {Theoretical} {Framework} for {Monetary} {Analysis}},
	volume = {78},
	number = {2},
	journal = {Journal of Political Economy},
	author = {Friedman, Milton},
	year = {1970},
	pages = {193--238},
}

@article{silva_modeling_2022,
	title = {Modeling supply-chain networks with firm-to-firm wire transfers},
	volume = {190},
	issn = {0957-4174},
	url = {https://www.sciencedirect.com/science/article/pii/S0957417421014846},
	doi = {10.1016/j.eswa.2021.116162},
	urldate = {2023-08-31},
	journal = {Expert Systems with Applications},
	author = {Silva, Thiago Christiano and Amancio, Diego Raphael and Tabak, Benjamin Miranda},
	month = mar,
	year = {2022},
	pages = {116162},
}

@article{aprigliano_using_2019,
	title = {Using {Payment} {System} {Data} to {Forecast} {Economic} {Activity}},
	volume = {15},
	language = {en},
	number = {4},
	journal = {International Journal of Central Banking},
	author = {Aprigliano, Valentina and Ardizzi, Guerino and Monteforte, Libero},
	year = {2019},
}

@techreport{buda_national_2022,
	title = {National {Accounts} in a {World} of {Naturally} {Occurring} {Data}: {A} {Proof} of {Concept} for {Consumption}},
	shorttitle = {National {Accounts} in a {World} of {Naturally} {Occurring} {Data}},
	url = {https://ideas.repec.org/p/cam/camjip/2220.html},
	language = {en},
	number = {2220},
	urldate = {2022-09-21},
	institution = {Faculty of Economics, University of Cambridge},
	author = {Buda, G. and Carvalho, V. M. and Hansen, S. and Mora, J. V. R. and Ortiz, Ã and Rodrigo, T.},
	month = jul,
	year = {2022},
	note = {Publication Title: Janeway Institute Working Papers},
}

@misc{hotte_national_2024,
	title = {National accounting from the bottom up using large-scale financial transactions data: {An} application to input-output tables},
	shorttitle = {National accounting from the bottom up using large-scale financial transactions data},
	url = {http://arxiv.org/abs/2407.14776},
	doi = {10.48550/arXiv.2407.14776},
	language = {en},
	urldate = {2025-01-27},
	publisher = {arXiv},
	author = {Hötte, Kerstin and Naddeo, Andreina},
	month = jul,
	year = {2024},
	note = {arXiv:2407.14776 [econ]},
}

@article{collibus_microvelocity_2025,
	title = {The microvelocity of money in {Ethereum}},
	volume = {14},
	copyright = {© The Author(s) 2025},
	issn = {2193-1127},
	url = {https://epjds.epj.org/articles/epjdata/abs/2025/01/13688_2024_Article_518/13688_2024_Article_518.html},
	doi = {10.1140/epjds/s13688-024-00518-6},
	language = {en},
	number = {1},
	urldate = {2025-02-21},
	journal = {EPJ Data Science},
	author = {Collibus, Francesco Maria De and Campajola, Carlo and Tessone, Claudio J.},
	month = dec,
	year = {2025},
	note = {Number: 1
Publisher: Springer Berlin Heidelberg},
	pages = {11},
}

@article{antras_measuring_2025,
	title = {Measuring the {Average} {Period} of {Production}},
	volume = {115},
	url = {https://antras.scholars.harvard.edu/sites/g/files/omnuum5876/files/2025-02/APP_PP_AEA_Submission_1_10_25.pdf},
	language = {en},
	journal = {AEA Papers and Proceedings},
	author = {Antràs, Pol and Tubdenov, Vitalii},
	month = may,
	year = {2025},
}

@article{guo_time-varying_2024,
	title = {A {Time}-{Varying} {Network} for {Cryptocurrencies}},
	volume = {42},
	issn = {0735-0015},
	url = {https://doi.org/10.1080/07350015.2022.2146695},
	doi = {10.1080/07350015.2022.2146695},
	number = {2},
	urldate = {2025-03-14},
	journal = {Journal of Business \& Economic Statistics},
	author = {Guo, Li and Härdle, Wolfgang Karl and Tao, Yubo},
	month = apr,
	year = {2024},
	note = {Publisher: ASA Website
\_eprint: https://doi.org/10.1080/07350015.2022.2146695},
	pages = {437--456},
}

@article{cramer_volume_1986,
	title = {The {Volume} of {Transactions} and the {Circulation} of {Money} in the {United} {States}, 1950–1979},
	volume = {4},
	issn = {0735-0015},
	url = {https://doi.org/10.1080/07350015.1986.10509517},
	doi = {10.1080/07350015.1986.10509517},
	number = {2},
	urldate = {2025-03-14},
	journal = {Journal of Business \& Economic Statistics},
	author = {Cramer, J. S.},
	month = apr,
	year = {1986},
	note = {Publisher: ASA Website
\_eprint: https://doi.org/10.1080/07350015.1986.10509517},
	pages = {225--232},
}

@article{serletis_random_1995,
	title = {Random {Walks}, {Breaking} {Trend} {Functions}, and the {Chaotic} {Structure} of the {Velocity} of {Money}},
	volume = {13},
	issn = {0735-0015},
	url = {https://www.tandfonline.com/doi/abs/10.1080/07350015.1995.10524619},
	doi = {10.1080/07350015.1995.10524619},
	number = {4},
	urldate = {2025-03-17},
	journal = {Journal of Business \& Economic Statistics},
	author = {Serletis, Apostolos},
	month = oct,
	year = {1995},
	note = {Publisher: ASA Website
\_eprint: https://www.tandfonline.com/doi/pdf/10.1080/07350015.1995.10524619},
	pages = {453--465},
}

@article{chen_large_2024,
	title = {Large {Spillover} {Networks} of {Nonstationary} {Systems}},
	volume = {42},
	issn = {0735-0015},
	url = {https://doi.org/10.1080/07350015.2022.2099870},
	doi = {10.1080/07350015.2022.2099870},
	number = {2},
	urldate = {2025-03-17},
	journal = {Journal of Business \& Economic Statistics},
	author = {Chen, Shi and Schienle, Melanie},
	month = apr,
	year = {2024},
	note = {Publisher: ASA Website
\_eprint: https://doi.org/10.1080/07350015.2022.2099870},
	pages = {422--436},
}

@article{ardakani_dynamics_2023,
	title = {The dynamics of money velocity},
	volume = {30},
	issn = {1350-4851},
	url = {https://doi.org/10.1080/13504851.2022.2083062},
	doi = {10.1080/13504851.2022.2083062},
	number = {13},
	urldate = {2025-03-17},
	journal = {Applied Economics Letters},
	author = {Ardakani, Omid M.},
	month = jul,
	year = {2023},
	note = {Publisher: Routledge
\_eprint: https://doi.org/10.1080/13504851.2022.2083062},
	pages = {1814--1822},
}

\section*{Appendix: Implementation details} \label{app:implementation}

Pairwise trajectory extraction is done using \texttt{follow-the-money}, an open-source piece of software available at \url{https://github.com/carolinamattsson/follow-the-money}~\parencite{mattsson_carolinamattssonfollow--money_2020}. The system specifications are noted in a configuration file included with the supplementary material. Briefly, the system boundary is precisely defined by the \textsc{disbursement} and \textsc{reclamation} transactions. The parameters used are noted in a script file included with the supplementary material. Tracing digital funds requires selecting an allocation heuristic, as discussed in Section~\ref{sec:method} and in~\textcite{mattsson_trajectories_2021}. We use the \texttt{--well-mixed} heuristic together with the \texttt{--pairwise} option to consider all possible pairs of sequential in- and out- transactions in all accounts. Fragments of received transactions with a size below the default \texttt{--size limit} 0.01 Sarafu are not tracked indefinitely; using the upper or lower bound for untracked durations gives the same estimates at the reported precision. Finally, a very small amount of Sarafu is mis-recorded in the data~\parencite{mattsson_sarafu_2022}. We employ the functionality provided in \texttt{follow-the-money} to infer the existence of missing funds, and find this to be an insignificant source of noise. 

The empirical distributions are produced using the held durations observed over the entire observation period (from 25 January 2020 through 15 June 2021). Weekly and monthly estimates are produced for the weeks and months that fall fully within this period, excluding the first week. Holding times that took place at \textit{system} accounts are filtered out. The total issued balance of the Sarafu system is computed at every hour as captured by the \textsc{disbursement} and \textsc{reclamation} transactions (see Section~\ref{sec:data}). The balance of \textit{system} accounts is subtracted from the total. Transaction volumes correspond to \textsc{standard} and \textsc{agent\_out} transactions (collectively referred to as \textit{transfers} throughout this work). 

Analysis is performed in \texttt{pandas}~\parencite{reback_pandas-devpandas_2020}, and figures are produced using \texttt{seaborn}~\parencite{waskom_seaborn_2021} and \texttt{matplotlib}~\parencite{caswell_matplotlibmatplotlib_2019}. Distributions are smoothed using kernel density estimation on a logarithmic scale, utilizing the \texttt{seaborn} default method for selecting the smoothing bandwidth.

\end{document}